%% file: main.tex
\documentclass{article}

    \PassOptionsToPackage{numbers, compress}{natbib}

    \usepackage[preprint]{neurips_2022}

\usepackage[utf8]{inputenc} 
\usepackage[T1]{fontenc}    
\usepackage{hyperref}       
\usepackage{url}            
\usepackage{booktabs}       
\usepackage{amsfonts}       
\usepackage{nicefrac}       
\usepackage{microtype}      
\usepackage{xcolor}         
\usepackage{graphicx}
\usepackage{tabularx}
\usepackage{adjustbox}
\newcolumntype{Y}{>{\centering\arraybackslash}X}
\usepackage{enumitem}

\title{STNDT: Modeling Neural Population Activity with a Spatiotemporal Transformer}

\author{%
  Trung Le \\
  Dept of Electrical and Computer Engineering\\
  University of Washington\\
  Seattle, WA 98195 \\
  \texttt{tle45@uw.edu} \\
   \And
   Eli Shlizerman \\
   Dept of Applied Mathematics \\
   Dept of Electrical and Computer Engineering \\
  University of Washington\\
  Seattle, WA 98195 \\
   \texttt{shlizee@uw.edu} \\
}

\begin{document}

\maketitle

\begin{abstract}
Modeling neural population dynamics underlying noisy single-trial spiking activities is essential for relating neural observation and behavior. A recent non-recurrent method - Neural Data Transformers (NDT) - has shown great success in capturing neural dynamics with low inference latency without an explicit dynamical model. However, NDT focuses on modeling the temporal evolution of the population activity while neglecting the rich covariation between individual neurons. In this paper we introduce SpatioTemporal Neural Data Transformer (STNDT), an NDT-based architecture that explicitly models responses of individual neurons in the population across 
time and space to uncover their underlying firing rates. In addition, we propose a contrastive learning loss that works in accordance with mask modeling objective to further improve the predictive performance. We show that our model achieves state-of-the-art performance
on ensemble level
in estimating neural activities across four neural datasets, demonstrating its capability to capture autonomous and non-autonomous dynamics spanning different cortical regions while being completely agnostic to the specific behaviors at hand. Furthermore, 
STNDT
spatial attention mechanism reveals consistently important subsets of neurons that play a vital role in driving the response of the entire population, providing interpretability and key insights into how the 
population of neurons
performs computation.
\end{abstract}

\input{introduction}
\input{methods}
\input{results}
\input{discussion}

\bibliographystyle{unsrt}

\input{main.bbl}
\end{document}

%% file: introduction.tex
\section{Introduction and Related Work}
One of the most prominent questions in systems neuroscience is how neurons perform computations that give rise to behaviors. 
Recent evidence suggests that computation in the brain could be governed at the population level \cite{yuste2015neuron, saxena2019towards}. Population of neurons are proposed
to obey an internal dynamical rule that drives their activities over time \cite{shenoy2013cortical, shenoy2021measurement}. Inferring these
dynamics on a single trial basis is crucial for understanding the relationship between neural population responses and behavior, subsequently
enabling the development of robust decoding schemes with wide applicability in 
brain-computer interfaces \cite{willett2021high, collinger2013high, jarosiewicz2015virtual}. However, modeling population dynamics on single trials is challenging due to the stochasticity of individual neurons making their spiking activity vary from trial to trial even when they are subject to identical stimuli.

A direct approach to reduce the trial-to-trial variability of neural responses could be to average responses over repeated trials of the same behavior \cite{levi2005role, nicolelis1995sensorimotor}, to convolve the neural response with a Gaussian kernel \cite{yu2008gaussian}, or in general, to define a variety of neural activity measures~\cite{shlizerman2012neural}. However, more success was found in approaches that explicitly model neural responses as a dynamical system, including methods treating the population dynamics as being linear \cite{kao2015single, gao2016linear}, switched linear \cite{linderman2017bayesian}, non-linear \cite{pandarinath2018inferring, ye2021representation}, or reduced projected nonlinear models~\cite{shlizerman2012neural}. Recent approaches leveraging recurrent neural networks (RNN) have shown 
promising progress
in modeling distinct components of a dynamical system - neural latent states, initial conditions and external inputs - on a moment-to-moment basis \cite{pandarinath2018inferring, keshtkaran2021large, zhu2021deep}. These sequential methods rely on continuous processing of neural inputs at successive timesteps, causing latency that hampers applicability in real-time decoding of neural signals. 
Consequently to RNN-based approaches,
Neural Data Transformer (NDT) \cite{ye2021representation} was proposed as a non-recurrent approach to improve inference speed by leveraging the transformers architecture which learns and predicts momentary inputs in parallel \cite{vaswani2017attention}. While successful,
NDT has only focused on modeling the relationship of 
neural population activity between timesteps while ignoring the rich covariation among individual neurons. Neurons in a population have been shown to have heterogeneous tuning profiles where each neuron has a different level of preference to a particular muscle movement direction~\cite{mahan2013motor, kohn2016correlations}. 
Neuron pairs also exhibit certain degree of correlation in terms of trial-to-trial variability (noise correlation) that affects the ability to decode the behaviors they represent~\cite{saxena2019towards, averbeck2006neural}. These spatial correlations characterize the amount of information that can be encoded in the neural population \cite{averbeck2006neural}, necessitating the need to model the neural population activity across both time and space dimensions.

In this work, we propose to incorporate the information distributed along the spatial dimension to improve the learning of neural population dynamics, and introduce \emph{SpatioTemporal}
Neural Data Transformer, 
an architecture based on Neural Data Transformer which explicitly learns both the spatial covariation between individual neurons and the temporal progression of the entire neural population. We summarize our main contributions as follows:

\begin{itemize}[leftmargin=*]
    \item We introduce Spatiotemporal Neural Data Transformer which allows the transformer to learn both the spatial coordination between neurons and the temporal progression of the population activity by letting neurons attend to each other while also attending over temporal instances.
    \item We propose a contrastive finetuning scheme, complementary to the mask modeling objective, to ensure the robustness of model prediction against induced noise augmentations.
    \item We validate our model's performance on four neural datasets in the publicly available Neural Latents Benchmark suite \cite{pei2021neural} and show that ensemble variants of our model
    outperforms other state-of-the-art methods, demonstrating its capability to model autonomous and non-autonomous neural dynamics in various brain regions while being agnostic to external behavior task structures.
    \item We show that the spatial attention, a feature unique to STNDT, identifies consistently important subsets of neurons that play an essential role in driving the response of the entire population. This exclusive attribute of STNDT provides interpretability and key insights into how the neural population distributes the computation workload among the neurons.
\end{itemize}

%% file: methods.tex
\section{Methods}

\begin{figure*}[t]
\begin{center}
    \includegraphics[width=0.9\linewidth]{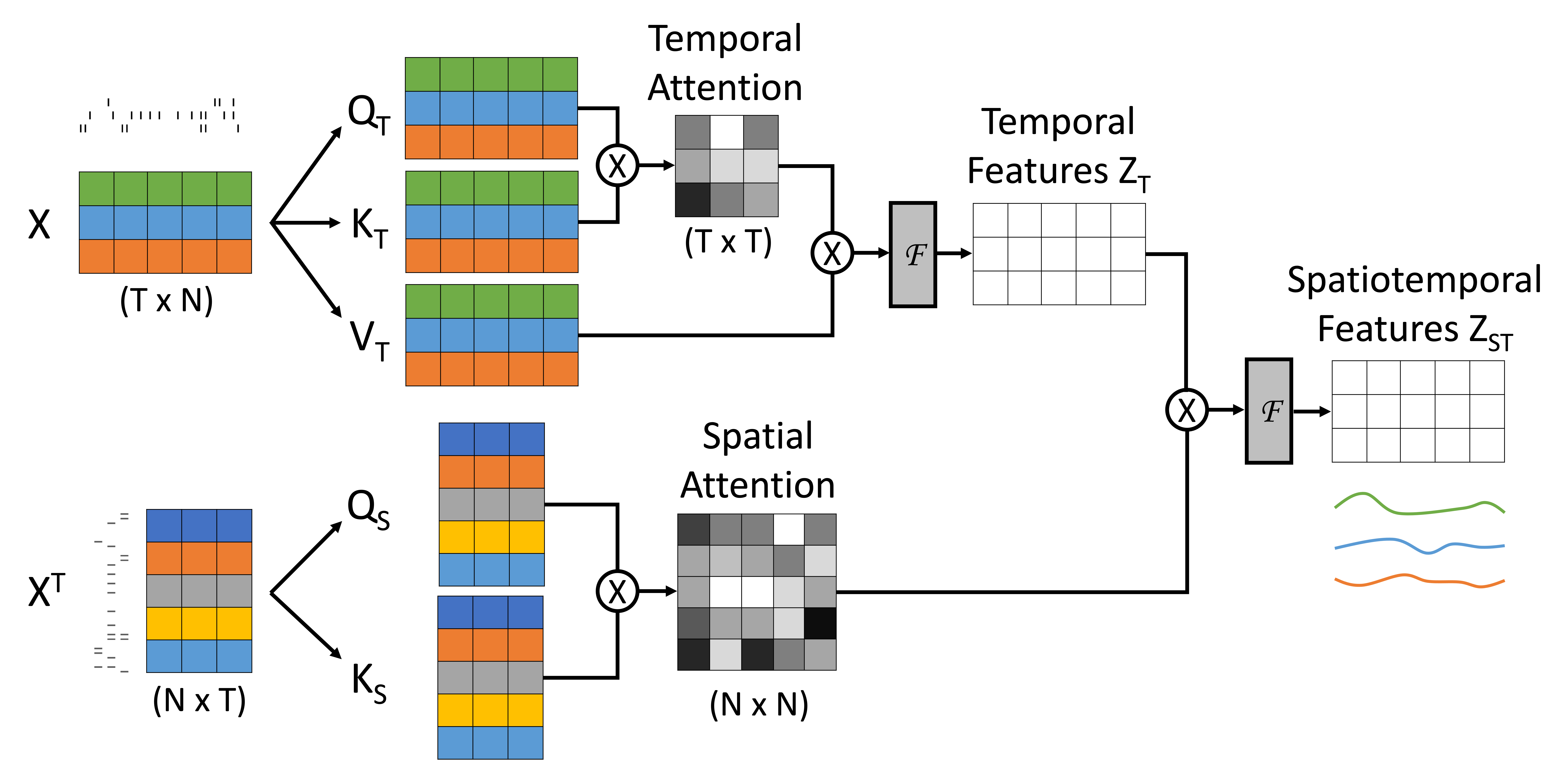}
\end{center}
   \caption{Spatiotemporal Neural Data Transformer (STNDT) architecture. Separate multihead self-attention modules are trained to learn spatial covariation and temporal progression of neural activities. Temporal attention feature matrix is treated as the matrix V upon which spatial attention is multiplied to give the final spatiotemporal features. The complete STNDT consists of multiple layers of such spatiotemporal attention modules.}
\label{fig:architecture}
\end{figure*}

\textbf{Problem formulation:}
Single-trial spiking activity of a neural population can be represented as a spatiotemporal matrix $X \in \mathbb{N}^{N \times T}$, where each row $X_i \in \mathbb{N}^T$ is the time series of one neuron, $N$ is the number of neurons in the population, and $T$ is the number of time bins for each trial. Each element $X_{nt}$ in the matrix is the number of action potentials (spikes) that neuron $n$ fires within the time bin $t$. Spike counts are assumed to be samples of an inhomogeneous Poisson process $P(\lambda(n,t))$ where $\lambda(n,t)$ is the underlying true firing rate of neuron $n$ at time $t$. The matrix $Y \in \mathbb{R}^{N \times T}$ containing $\lambda(n,t)$ fully represents the dynamics of the neural population and explains the observable spiking data of the respective trial. 
We propose to learn the mapping $\phi(X;W): X \rightarrow Y$ by the Spatiotemporal Transformer with the set of weights $W$. 

\textbf{Spatiotemporal Neural Data Transformer:}
At the core of the transformer architecture is the multihead attention mechanism, where feature vectors learn to calibrate the influence of other feature vectors in their transformation. Spike trains are embedded into a feature matrix $\tilde{X}$ with added positional encoding to preserve order information as initially proposed in \cite{vaswani2017attention}.

A set of three matrices $W^Q$, $W^K$, $W^V$ $\in \mathbb{R}^{N \times N}$ are learned to transform $T$ $N$-dimensional embedding $\tilde{X}=\{\tilde{x}_1, \tilde{x}_2, ..., \tilde{x}_T\}$ to queries $Q=\tilde{X}W^Q$, keys $K=\tilde{X}W^K$ and values $V=\tilde{X}W^V$, upon which latent variable $Z$ is computed as:
\begin{equation}\label{eq:1}
    Z = \textnormal{Attention}(Q,K,V) = \mathcal{F}\left(\textnormal{softmax}\left(\frac{QK^\top}{\sqrt{N}}\right)V\right)
\end{equation}
The outer product of $QK^T$ represents the attention each $x_i$ pays to all other $x_j$ and determines how much influence their values $v_j$ have on its latent output $z_i$. $\mathcal{F}$ is the sequence of concatenating multiple heads and feeding through a feedforward network with ReLU activation \cite{vaswani2017attention}.

Implementations of transformers in popular applications such as in natural language processing literature consider each feature vector $x_i$ as an $N$-dimensional token in a sequence, equivalent to a word in a sentence. Elements in the $N$-dimensional vector therefore serve as a convenient numerical representation and do not have inherent relationships among them. The attention mechanism thus only models the relationship between tokens in a sequence. In our application, each feature vector $x_i$ is a collection of firing activities of $N$ physical neurons among which there exists an interrelation as neuronal population acts as a coordinated structure with complex interdependencies rather than standalone individuals. We therefore propose to model both the temporal relationship - the evolution of neural activities - and the spatial relationship - covariability of neurons - by learning two separate multihead attention blocks (Figure \ref{fig:architecture}). The temporal latent state $Z_\mathcal{T}$ is computed with temporal attention block as in Equation \ref{eq:1}. In parallel, spatial attention block operates on the transposed embedding $\tilde{X}^\top$ and learns an attention weights matrix signifying the relationship between neurons:
\begin{equation}
    A_\mathcal{S} = \textnormal{softmax}\left(\frac{Q_\mathcal{S}K_\mathcal{S}^\top}{\sqrt{T}}\right)
\end{equation}
where $Q_\mathcal{S} = \tilde{X}^\top W_\mathcal{S}^Q$ and $K_\mathcal{S} = \tilde{X}^\top W_\mathcal{S}^K$. 

This $A_\mathcal{S}$ matrix is then multiplied with the temporal latent state $Z_\mathcal{T}$ to incorporate the influence of spatial attention on the final spatiotemporal latent state $Z_\mathcal{ST}$:
\begin{equation}
    Z_\mathcal{ST} = \mathcal{F}(A_\mathcal{S}Z_\mathcal{T})
\end{equation}
\textbf{Mask modeling and contrastive losses:}
Similar to \cite{ye2021representation}, we train the spatiotemporal transformer in an unsupervised way with BERT's mask modeling objective \cite{devlin2018bert}. During training, a random subset of spike bins along both spatial and temporal axes of input $X$ are masked (zero-ed out or altered) and the transformer is asked to reconstruct the log firing rate at the masked bins such that the Poisson negative log likelihood is minimized:
\begin{equation}\label{eq:mask_loss}
    \mathcal{L}_{mask} = \sum_{i=1}^{N}\sum_{j=1}^{T}{\textnormal{exp}(\tilde{z}_{ij}) - \tilde{x}_{ij} \tilde{z}_{ij}}
\end{equation}
where $\tilde{z}_{ij}$ and $\tilde{x}_{ij}$ are the output firing rate and input spike of neuron $i$ at timestep $j$ if location $ij$ is masked.

Neural dynamics are shown to be embedded in a low-dimensional space, i.e. model prediction should be fairly consistent when a smaller subset of neurons are used compared to when the entire population is taken into account. Furthermore, in stereotyped behaviors often found in neuroscience experiments, trials with the same condition should yield similar output firing rate profiles. Therefore, to enhance robustness of model prediction to neural firing variability, we further constrain model firing rate outputs by a contrastive loss, such that different augmentations of the same trial input remain closer to each other and stay distant to other trial inputs. We adopt the NT-XEnt contrastive loss 
introduced in 
\cite{chen2020simple}:
\begin{equation}
    \mathcal{L}_{contrastive} = \sum_{ij}{l_{ij}} = \sum_{ij}{-\textnormal{log} \frac{\textnormal{exp}(\textnormal{sim}(z_i, z_j)/\tau)}{\sum_{k=1}^{2N} \mathbf{1}_{k \ne i} \textnormal{exp}(\textnormal{sim}(z_i, z_k)/\tau) }}
\end{equation}
where $\textnormal{sim}(u, v) = u\top v /(\|u\|\|v\|)$ is the cosine similarity between two predictions $u$ and $v$ on two different augmentations of input $x$ and $\tau$ is the temperature parameter. 

We define the augmentation transformation as random dropout and alteration of spike counts on the original input matrix $X$. We first train the transformer with mask modeling loss in Eq.~\ref{eq:mask_loss} and finetune it with the addition of contrastive loss as we found that applying contrastive loss at this late stage when predictions are pretty stable would bring the most improvement to the model performance. 

\textbf{Bayesian hyperparameter tuning:}
We follow \cite{aestudio} to use Bayesian optimization for hyperparameters tuning. We observe that the primary metrics co-smoothing bits/spike (co-bps) are not well correlated with the mask loss (see Figure \ref{fig:metrics}), while co-bps, vel $R^2$, psth $R^2$ and fp-bps are more pairwise correlated.  Therefore, we run Bayesian optimization to optimize co-bps for $M$ models then select the best $N$ models as ranked by validation co-bps, and ensemble them by taking the mean of the predicted rates of these $N$ models.

%% file: results.tex
\section{Experiments and results}
\begin{figure}[t]
\begin{center}
  \includegraphics[width=0.9\linewidth]{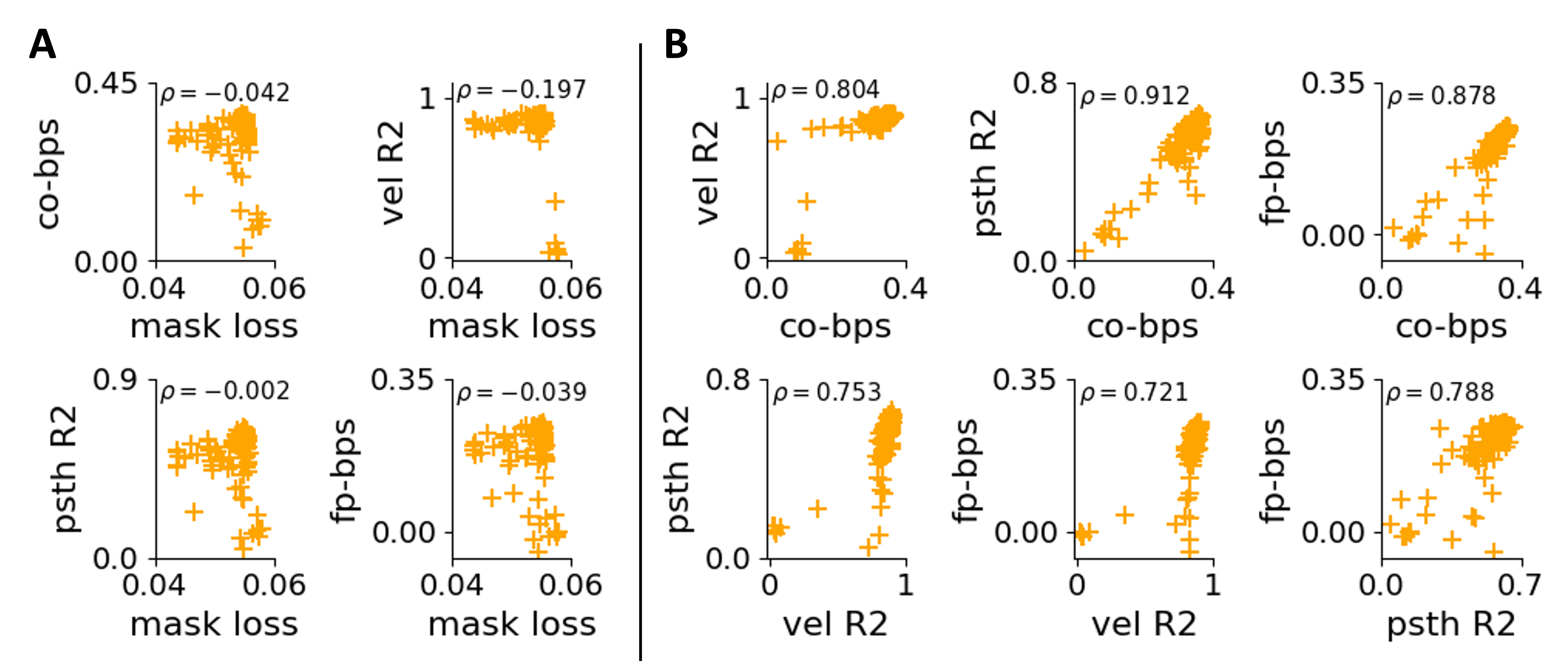}
\end{center}
  \caption{Correlations of evaluation metrics. \textbf{A}: Four evaluation metrics of 120 models obtained from Bayesian hyperparameter optimization on MC\_Maze dataset are plotted against mask loss. The evaluation metrics do not correlate well with mask loss. \textbf{B}: The four metrics are more correlated with each other, therefore we opted for co-bps as the objective for Bayesian hyperparameter optimization.}
\label{fig:metrics}
\end{figure}
\textbf{Datasets and evaluation metrics:}
We evaluate our model performance on four neural datasets in the publicly available Neural Latents Benchmark \cite{pei2021neural}: MC\_Maze, MC\_RTT, Area2\_Bump, and DMFC\_RSG. The 4 datasets cover autonomous and non-autonomous neural population dynamics recorded on rhesus macaques in a variety of behavioral tasks (delayed reaching, self-paced reaching, reaching with perturbation, time interval reproduction) spanning multiple brain regions (primary motor cortex, dorsal premotor cortex, somatosensory cortex, dorso-medial frontal cortex). The diverse scenarios and systems  offer comprehensive evaluation of a latent variable model and serve as a standardized benchmark for comparison between different modeling approaches.
We use different metrics to measure performance of our model depending on the particular behavior task of each dataset, following the standard evaluation pipeline in \cite{pei2021neural}. We evaluate and report our model performance on the hidden test split held by NLB to have a fair comparison with other state-of-the-art (SOTA) methods. See \cite{pei2021neural} for further details of evaluation strategy and how the metrics are calculated.
\begin{itemize}[leftmargin=*]
\item \textbf{Co-smoothing (co-bps)}: the primary metric, measuring the ability of the model to predict activity of held-out neurons it has not seen during training.

\item\textbf{Behavior decoding (vel $\mathbf{R^2}$ or tp-corr)}: measures how useful the model firing rates prediction can be used to decode behavior (the velocity of primate's hand in the cases of MC\_Maze and Areas\_Bump datasets, or the correlation between neural speed and time between Set cue and Go response in DMFC\_RSG dataset).

\item\textbf{Match to peri-stimulus time histogram (psth $\mathbf{R^2}$)}: indicates how well predicted firing rates match the peri-stimuls time histogram in repeated, stereotyped task structures.

\item\textbf{Forward prediction (fp-bps)}: measures model's ability to predict unseen future activity of the neural population.
\end{itemize}

\textbf{Baselines:}
We compare STNDT against the following baselines, all of which have been evaluated using the same held-out test split.
\begin{itemize}[leftmargin=*]
\item\textbf{Smoothing} \cite{pei2021neural}: A simple method where a Gaussian kernel is convolved with held-in spikes to produce smoothed held-in firing rates. Then a Poisson Generalized Linear Model (Poisson GLM) is fitted from the held-in smoothed rates to held-out rates.

\item\textbf{GPFA} \cite{yu2008gaussian}: extracts population latent states as a smooth and low dimensional evolution by combining smoothing and dimension reduction in a common probabilistic framework.

\item\textbf{SLDS} \cite{linderman2017bayesian}: models neural dynamics as a switching linear dynamical system, which breaks down nonlinear data into sequences of simpler dynamical modes.

\item\textbf{AutoLFADS} \cite{keshtkaran2021large}: models population activity as a non-linear dynamical system with  bi-directional recurrent neural networks at the core and an automatic, scalable framework of hyperparameter tuning.

\item\textbf{MINT} \cite{mint}: an interpretable decode algorithm that exploits the sparsity and stereotypy of neural activity to interpolate neural states using a library of canonical neural trajectories.

\item\textbf{iLQR-VAE} \cite{schimel2021ilqr}: improves upon LFADS with iterative linear quadratic regulator algorithm, an optimization-based recognition model to replace RNN as the inference network.

\item\textbf{NDT} \cite{ye2021representation}:
leverages transformer architecture with some adaption to neural data to model temporal progression of neural activity across time. AESMTE1 is the best single model and AESMTE3 is the best emsemble of multiple models found as a result of Bayesian hyperparameter tuning \cite{aestudio}. 
\end{itemize}
\begin{figure*}[t]
\begin{center}
  \includegraphics[width=0.85\linewidth]{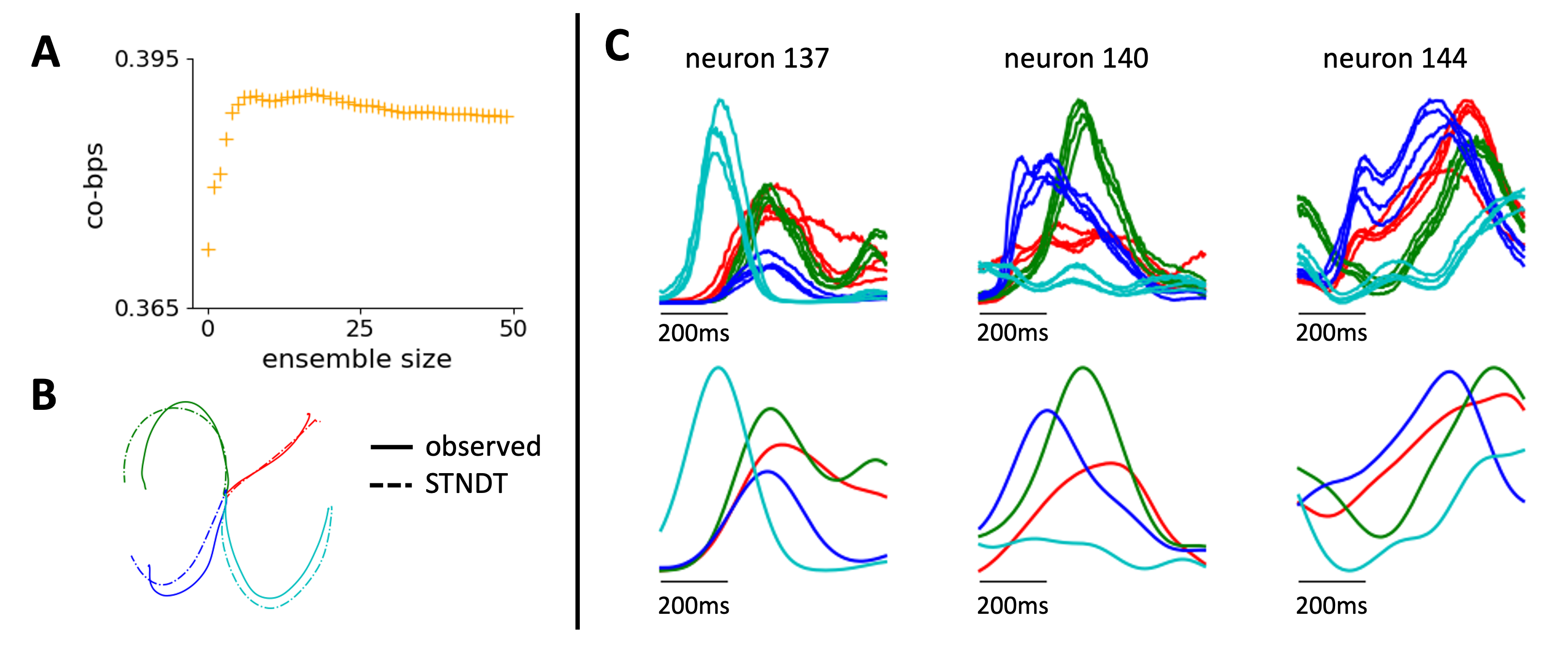}
\end{center}
  \caption{\textbf{A}: co-bps metrics improves when multiple models are ensembled together. \textbf{B}: STNDT facilitates accurate inference of behavior from spiking data. Decoded hand trajectories from 4 trials (dashed line) closely match the ground truth trajectories (solid line). \textbf{C}: STNDT uncovers the stereotyped feature of neural activity in structured behaviors. Firing rate prediction and PSTHs of three example neurons are shown. Trials belonging to the same condition are plotted with the same color (4 trials per condition shown). All results are shown for MC\_Maze dataset.}
\label{fig:ensemble_psth}
\end{figure*}
\subsection{Spatiotemporal transformer achieves state-of-the-art performance in modeling autonomous dynamics}
We first tested STNDT on recordings of dorsal premotor (PMd) and motor cortex (M1) of a monkey performing a delayed reaching task (MC\_Maze dataset) to evaluate the ability of STNDT to uncover single-trial population dynamics in a highly structured, stereotyped behavior. The dataset has been studied extensively in previous work \cite{pandarinath2018inferring, keshtkaran2021large, ye2021representation} and presents a unique opportunity for us to compare our method with other state-of-the-art approaches. The dataset consists of 2869 trials of monkey performing a center-out reaching task in a maze with obstructing barriers, composing 108 different conditions for straight and curved reaching trajectories. The monkey is trained to hold the cursor at the center while the target is presented and only move the cursor to reach the target after a `Go' cue. The neural dynamics during the preparation and execution periods is well modeled as an autonomous dynamical system \cite{pandarinath2018inferring}. 

We observed that by explicitly modeling spatial attention, STNDT outperformed other state-of-the-art methods and improved the original NDT's ability to model autonomous single-trial dynamics as measured by the negative log likelihood of unobserved neural activity. The single STNDT model improved both Poisson log likelihood of heldout neurons (co-bps) and heldout timesteps (fp-bps). The performance is further increased by aggregating multiple STNDT models, achieving score of $0.3862$ and $0.2686$ on co-bps and fp-bps respectively, compared to $0.3676$ and $0.2589$ of current state-of-the-art AESMTE3 \ref{table:performance1}.

Since MC\_Maze features repeated trials, the prediction of any latent variable models should uncover stereotypical patterns of neuronal responses for trials belonging to the same conditions. Therefore, we computed peri-stimulus time histogram (PSTH) which is the average of neural population response across trials of the same condition, and measure the $R^2$ matching of model prediction to this PSTH. We observed that with the help of spatial modeling and contrastive loss, STNDT boosts NDT ability to recover this stereotyped firing pattern, reaching $0.6693$ compared to $0.6683$ of ensemble NDT (AESMTE3). We show in Figure \ref{fig:ensemble_psth} several responses of example neurons. STNDT firing rates prediction of trials under the same condition exhibit a consistent, stable PSTH as desired.
\subsection{Spatiotemporal transformer enhances prediction of neural population activity during cognitive task}
Dorsomedial frontal cortex (DMFC) is believed to serve as an intermediate layer between low-level sensory and motor areas, and possess distinct confluence of internal dynamics and inputs \cite{rigotti2013importance, sohn2019bayesian}. We are therefore interested to see if characterizing spatial relationship alongside temporal relationship and incorporating contrastive loss could help STNDT better model the dynamics in this brain region. We tested STNDT on the DMFC\_RSG dataset \cite{pei2021neural} consisting of recordings from a rhesus macaque performing a time-interval reproduction task. The monkey is presented two `Ready' and `Set' stimuli separated by a specific time interval $t_s$ while fixating eye and hold the joystick at the center position. It then has to execute a `Go' response by either an eye saccade or joystick movement such that the time interval $t_p$ between its reponse and the `Set' cue is sufficiently close to $t_s$. 

We observed that STNDT improves NDT's performance on both single and ensemble level. STNDT achieves $0.1859$ and $0.1940$ co-bps score, outperform NDT's $0.1733$ and $0.1886$ for single and ensemble models, respectively. STNDT also enhances NDT's ability to model future neural activity, as it boosts NDT's fp-bps scores of $0.1511$ to $0.1601$ for single best model, and $0.1828$ to $0.1910$ for the ensemble of models. STNDT also uncovers stereotyped features that are consistent across repeated trials of the same behavior conditions, as measured by the match to peri-stimulus time histogram (PSTH $R^2$). STNDT outperforms NDT on single best model ($0.6051$ compared to $0.5267$) as well as on the ensemble of models ($0.6452$ compared to $0.6064$), only seconded by MINT whose model optimizes the preservation of neural trajectories across trials and hence intuitively would achieve a high score in this criterion.
\begin{table*}[t!]
  \caption{Performance of STNDT as compared to SOTA methods on MC\_Maze and MC\_RTT datasets}
  \label{table:performance1}
  \centering
  \small
  \begin{tabularx}{\textwidth}{@{}cYYcYYYY@{}}
    \toprule
    & \multicolumn{4}{c}{MC\_Maze} & \multicolumn{3}{c}{MC\_RTT} \\
    \cmidrule{2-8}
    Methods     & co-bps$\uparrow$     & vel $R^2$$\uparrow$   & psth $R^2$$\uparrow$   & fp-bps$\uparrow$  & co-bps$\uparrow$     & vel $R^2$$\uparrow$ & fp-bps$\uparrow$ \\
    \midrule
    GPFA     & $0.1872$       & $0.6399$  & $0.5150$  & $-$ & $0.1548$  & $0.5339$ & $-$   \\
    Smoothing    & $0.2109$ & $0.6238$  & $0.1853$  & $-$ & $0.1468$  & $0.4142$ & $-$  \\
    SLDS    & $0.2249$       & $0.7947$  & $0.5330$  & $1.1579$ & $0.1649$  & $0.5206$ & $0.0620$  \\
    MINT    & $0.3304$       & $\mathbf{0.9121}$  & $\mathbf{0.7496}$  & $0.2076$ & $0.1676$  & $0.5953$ & $0.1012$ \\
    AutoLFADS     & $0.3364$ & $0.9097$  & $0.6360$  & $0.2349$ & $0.1868$  & $0.6167$ & $0.1213$  \\
    iLQR-VAE     & $0.3559$ & $0.8840$  & $0.6062$  & $0.1480$ & $-$ & $-$ & $-$  \\
    AESMTE1 (single)     & $0.3599$       & $0.9105$  & $0.6641$  & $0.2470$ & $0.1927$  & $\mathbf{0.6627}$ & $0.1229$  \\
    AESMTE3 (ensemble) & $0.3676$  & $0.9114$ & $0.6683$  & $0.2589$ & $0.2053$  & $0.6334$ & $\mathbf{0.1344}$  \\
    \midrule
    STNDT single (ours) & $0.3691$  & $0.8985$ & $0.6567$  & $0.2505$ & $0.1865$  & $0.5988$ & $0.0964$  \\
    STNDT ensemble (ours) & $\mathbf{0.3862}$  & $0.9095$ & $0.6693$  & $\mathbf{0.2686}$ & $\mathbf{0.2095}$  & $0.6270$ & $0.1244$  \\
    \bottomrule
  \end{tabularx}
\end{table*}
\begin{table*}[t!]
  \caption{Performance of STNDT as compared to SOTA methods on Area2\_Bump and DMFC\_RSG datasets}
  \label{table:performance2}
  \centering
  \small
  \begin{tabularx}{\textwidth}{@{}cYYYYYcYr@{}}
    \toprule
    & \multicolumn{4}{c}{Area2\_Bump} & \multicolumn{4}{c}{DMFC\_RSG} \\
    \cmidrule{2-9}
    Methods  & co-bps$\uparrow$     & vel $R^2$$\uparrow$   & psth $R^2$$\uparrow$   & fp-bps$\uparrow$ & co-bps$\uparrow$ & tp-corr$\downarrow$   & psth $R^2$$\uparrow$ & fp-bps$\uparrow$ \\
    \midrule
    GPFA      & $0.1680$ & $0.5975$  & $0.5289$ & $-$  & $0.1176$ & $-0.3763$  & $0.2142$ & $-$     \\
    Smoothing     & $0.1544$ & $0.5736$  & $0.2084$ & $-$  & $0.1202$ & $-0.5139$  & $0.2993$ & $-$     \\
    SLDS    & $0.1960$ & $0.7385$  & $0.5740$ & $0.0242$  & $0.1243$ & $-0.5412$  & $0.3372$ & $-0.0418$     \\
    MINT      & $0.2735$ & $0.8877$  & $\mathbf{0.9135}$ & $0.1483$  & $0.1821$ & $-0.6929$  & $\mathbf{0.7013}$ & $0.1650$     \\
    AutoLFADS      & $0.2569$ & $0.8492$  & $0.6318$ & $0.1505$  & $0.1829$ & $\mathbf{-0.8248}$  & $0.6359$ & $0.1844$     \\
    iLQR-VAE       & $-$ & $-$  & $-$ & $-$  & $-$ & $-$  & $-$ & $-$     \\
    AESMTE1 (single)   & $0.2801$ & $0.8675$  & $0.6367$ & $0.1523$  & $0.1733$ & $-0.6189$  & $0.5267$ & $0.1511$     \\
    AESMTE3 (ensemble)  & $0.2860$ & $\mathbf{0.8999}$  & $0.7109$ & $\mathbf{0.1603}$  & $0.1886$ & $-0.7601$  & $0.6064$ & $0.1828$     \\
    \midrule
    STNDT single (ours)   & $0.2818$ & $0.8766$  & $0.6454$ & $0.1357$  & $0.1859$ & $-0.5205$  & $0.6051$ & $0.1601$\\
    STNDT ensemble (ours)   & $\mathbf{0.2898}$ & $0.8913$  & $0.7368$ & $0.1476$  & $\mathbf{0.1940}$ & $-0.4857$  & $0.6452$ & $\mathbf{0.1910}$\\
    \bottomrule
  \end{tabularx}
\end{table*}
\subsection{Spatial attention mechanism identifies important subsets of neurons driving the population dynamics}
\begin{figure*}[t]
\begin{center}
    \includegraphics[width=0.95\linewidth]{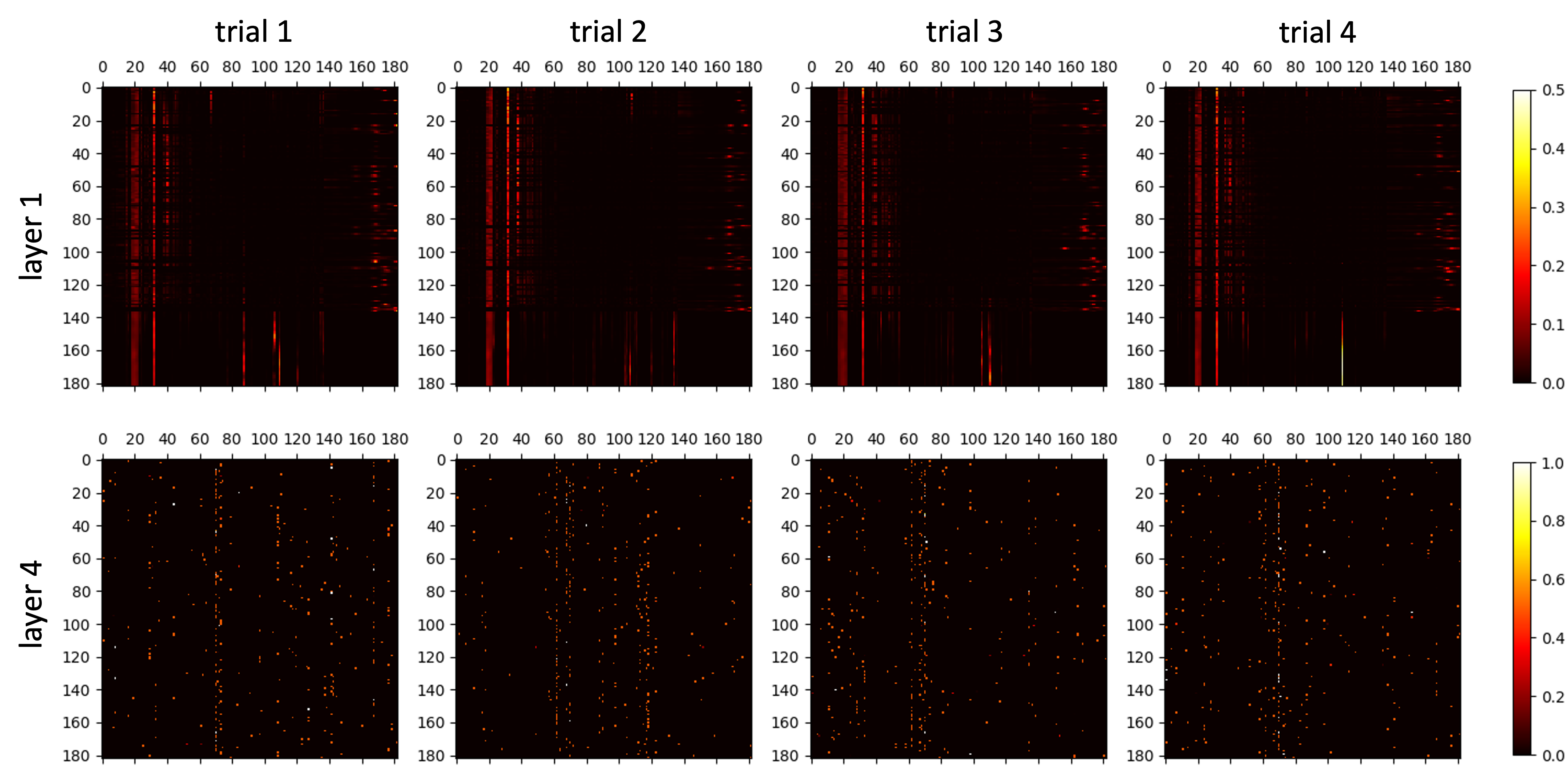}
\end{center}
   \caption{Visualization of STNDT's spatial attention weights in layer 1 and layer 4 of four example trials. Attention weights in layer 1 reveal a consistent subset of neurons that are heavily attended to by all neurons in the population. The attention becomes more dispersed in deeper layers. Results are shown for 182 neurons in MC\_Maze dataset.}
\label{fig:attention_weights}
\end{figure*}

\begin{figure*}[ht]
\begin{center}
    \includegraphics[width=0.8\linewidth]{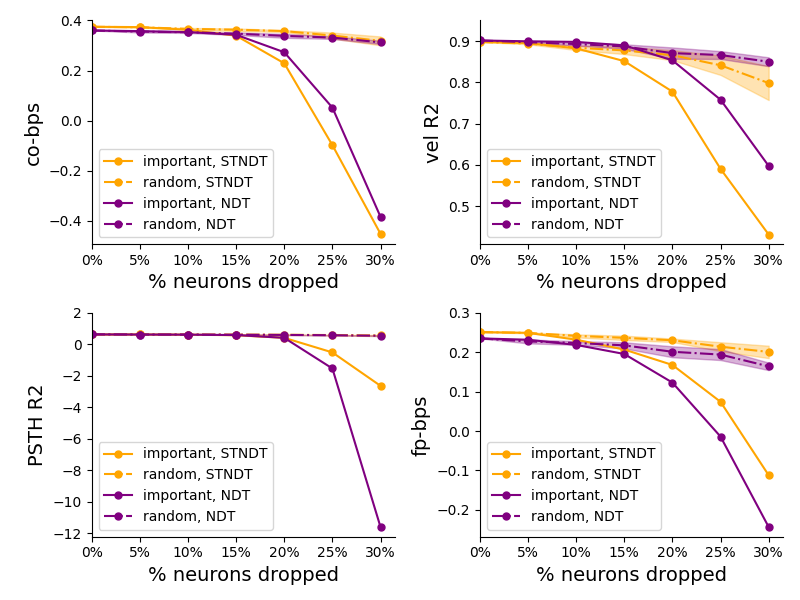}
\end{center}
   \caption{
Spatial attention module, unique to STNDT, identifies important neurons that are the main driving force of population response to behavioral task.
Performance of STNDT as measured by four evaluation metrics are plotted as neurons are incrementally dropped from input neural population. Performance significantly deteriorates when important neurons identified by STNDT are dropped, while only decreases slightly when random neurons are dropped. The effect of important neurons indentified by STNDT generalizes to vanilla NDT, which lacks a spatial attention structure. Shaded region represents 2 standard error of the mean. Results are shown for MC\_Maze dataset.}
\label{fig:dropout}
\end{figure*}

The weights of attention matrix have been used as a tool to provide certain level of interpretability for attention-based models \cite{clark2019does, kovaleva2019revealing, lin2019open, ghader2017does}. The interpretability is built upon the fact that attention weights signify how much influence other inputs have on a particular input in deciding its final outcome. This influence might align with some human interpretable meaning, such as linguistic patterns \cite{reif2019visualizing}. In Figure \ref{fig:attention_weights}, we visualize spatial attention weights obtained from STNDT on the MC\_Maze dataset across 4 attention layers.
Interestingly, spatial attention shows that in early layers, only a small subsets of neurons in the population are consistently attended to by all neurons. The spatial attention tends to disperse as the model goes to deeper layers. Strikingly, the subset of heavily-attended neurons stays relatively identical across different trials, hinting that these neurons might play a crucial role in driving the population response to the behavior task. We further tested this hypothesis by incrementally dropping the neurons heavily attended to (i.e. zeroing out their spiking activity input to the model) in a descending order of their attention weights identified in the first layer. We observed that dropping these important neurons identified by STNDT caused a significant decline in the model performance (Figure \ref{fig:dropout}). The performance decline was significantly more than the case where the same number of random neurons are dropped. 
To rule out the possible case that dropping neurons only has adverse effect on the spatial attention module but that effect propagates to the subsequent modules and indirectly impacts the performance of the overall STNDT pipeline, we repeated the experiment on the vanilla NDT model which, unlike STNDT, lacks a spatial attention structure. Interestingly, we observed the same performance deterioration when we dropped the spiking activity of STNDT-identified important neurons and asked
a pretrained
vanilla NDT
to make inference on the resulting inputs.
This finding suggests that the impact of the 
important neurons that only STNDT can identify
might potentially generalize
to other latent variable models that without input from these neurons, some latent variable models might not function optimally. 
\subsection{Ablation Study: Contrastive loss encourages consistency of model prediction and improves performance}
We conduct an ablation study to assess the effectiveness of contrastive loss on the overall performance of STNDT. Tables \ref{table:ablation1} and \ref{table:ablation2} report how the model scores on different metrics across all four datasets on the single and ensemble levels. In general, we observe that having contrastive loss further improves the performance of STNDT on predicting neural activity of heldout neurons (co-bps) and heldout timesteps (fp-bps). The contribution of contrastive loss is most eminent on MC\_Maze dataset.
\begin{table*}[t!]
  \caption{Ablation Study: Performance of STNDT on MC\_Maze and MC\_RTT datasets with and without contrastive loss (CL) on single and ensemble levels.}
  \label{table:ablation1}
  \centering
  \small
  \begin{tabularx}{\textwidth}{@{}cYYcYYYr@{}}
    \toprule
    & \multicolumn{4}{c}{MC\_Maze} & \multicolumn{3}{c}{MC\_RTT} \\
    \cmidrule{2-8}
    \multicolumn{1}{c}{Methods}  & co-bps$\uparrow$     & vel $R^2$$\uparrow$   & psth $R^2$$\uparrow$   & fp-bps$\uparrow$ & co-bps$\uparrow$ & vel $R^2$$\uparrow$ & fp-bps$\uparrow$ \\
    \midrule
    AESMTE1 (single)     & $0.3599$       & $0.9105$  & $0.6641$  & $0.2470$ & $0.1927$  & $0.6627$ & $0.1229$  \\
    AESMTE3 (ensemble) & $0.3676$  & $0.9114$ & $0.6683$  & $0.2589$ & $0.2053$  & $0.6334$ & $0.1344$  \\
    \midrule
    STNDT single w/o CL      & $0.3668$ & $0.8979$  & $0.6549$ & $0.2471$  & $0.1865$ & $0.5988$  & $0.0964$    \\
    STNDT single w/ CL     & $0.3691$ & $0.8985$  & $0.6567$ & $0.2505$  & $0.1865$ & $0.5988$  & $0.0964$ \\
    STNDT ensemble w/o CL    & $0.3843$ & $0.9090$  & $0.6686$ & $0.2675$  & $0.2065$ & $0.6352$  & $0.1260$     \\
    STNDT ensemble w/ CL   & $0.3862$ & $0.9095$  & $0.6693$ & $0.2686$  & $0.2095$ & $0.6270$  & $0.1244$ \\
    \bottomrule
  \end{tabularx}
\end{table*}
\begin{table*}[t!]
  \caption{Ablation Study: Performance of STNDT on Area2\_Bump and DMFC\_RSG datasets with and without contrastive loss (CL) on single and ensemble levels.}
  \label{table:ablation2}
  \centering
  \small
  \begin{tabularx}{\textwidth}{@{}cYYYYYYYY@{}}
    \toprule
    & \multicolumn{4}{c}{Area2\_Bump} & \multicolumn{4}{c}{DMFC\_RSG} \\
    \cmidrule{2-9}
    \multicolumn{1}{c}{Methods}  & co-bps$\uparrow$     & vel $R^2$$\uparrow$   & psth $R^2$$\uparrow$   & fp-bps$\uparrow$ & co-bps$\uparrow$ & tp-corr$\downarrow$   & psth $R^2$$\uparrow$ & fp-bps$\uparrow$ \\
    \midrule
    AESMTE1 (single)   & $0.2801$ & $0.8675$  & $0.6367$ & $0.1523$  & $0.1733$ & $-0.6189$  & $0.5267$ & $0.1511$     \\
    AESMTE3 (ensemble)  & $0.2860$ & $0.8999$  & $0.7109$ & $0.1603$  & $0.1886$ & $-0.7601$  & $0.6064$ & $0.1828$     \\
    \midrule
    STNDT single w/o CL      & $0.2765$ & $0.8773$  & $0.7169$ & $0.1498$ & $0.1824$  & $-0.5059$ & $0.6134$  & $0.1473$     \\
    STNDT single w/ CL     & $0.2818$ & $0.8766$  & $0.6454$ & $0.1357$  & $0.1859$ & $-0.5205$  & $0.6051$ & $0.1601$ \\
    STNDT ensemble w/o CL    & $0.2904$ & $0.8937$  & $0.7303$ & $0.1491$  & $0.1931$ & $-0.5186$  & $0.6429$ & $0.1888$     \\
    STNDT ensemble w/ CL   & $0.2898$ & $0.8913$  & $0.7368$ & $0.1476$  & $0.1940$ & $-0.4857$  & $0.6452$ & $0.1910$     \\
    \bottomrule
  \end{tabularx}
\end{table*}

%% file: discussion.tex
\section{Discussion}\label{sec:discussion}
In this paper we presented Spatiotemporal Neural Data Transformer, a novel architecture based upon Neural Data Transformer \cite{ye2021representation} that explicitly learns the covariation among individual neurons in the population alongside the momentary evolution of the population spiking activity in order to infer the underlying firing rates behind highly variable single-trial spike trains. By incorporating self-attention along both spatial and temporal dimensions, as well as a contrastive loss, STNDT enhances NDT's ability to model dynamics spanning a variety of tasks and brain regions as measured by the accurate prediction of activity of unseen neurons and timesteps, as well as the discovery of stereotyped features across trials of the same behavior conditions. STNDT also maintains a comparable ability as NDT to allow decent decoding of behavior from its rate prediction. Finally, the novel spatial attention mechanism unique to STNDT brings about valuable interpretability as it discovers influential subsets of neurons whose activities contain salient information about the response of the entire neural population without which some latent variable models might not function optimally. 

Although STNDT with contrastive loss has demonstrated on both single and ensemble levels great success in modeling autonomous dynamics in premotor and primary motor cortices (MC\_Maze) and non-autonomous dynamics in dorsomedial frontal cortex during cognitive task (DMFC\_RSG), we have not observed 
that the incorporation of spatial attention to STNDT without contrastive loss brought about an improvement in the primary metric co-bps on the single model level for datasets with non-autonomous dynamics in small scale (Area2\_Bump) and unstructured behavior (MC\_RTT) (Tables \ref{table:ablation1} and \ref{table:ablation2}). We hypothesize that the approach of artificially splitting continuous data into overlapping "trials" in MC\_RTT and the relatively small scale of Area2\_Bump potentially hinder the effective learning of spatial attention features, since the codependence of neurons' firing rates might be best expressed and identified with sufficiently large recordings and well structured behavior task design, which were the case in MC\_Maze and DFMC\_RSG datasets. 